\documentclass[twocolumn,amsmath,amssymb]{article}

\usepackage{graphicx}
\usepackage{dcolumn}
\usepackage{bm}

\title{Temperature dependence of exciton recombination in semiconducting
single-wall carbon nanotubes}

\author{S. Berger \and C. Voisin \and G. Cassabois \and C. Delalande \and P.
Roussignol \\ 
Laboratoire Pierre Aigrain, \'Ecole
Normale Sup\'erieure\\ 24, rue Lhomond, 75005 Paris, France \\
\and X. Marie \\
Laboratoire de Nanophysique, Magn\'etisme et Opto\'electronique,
INSA \\ 135 avenue de Rangueil, 31077 Toulouse, France}

\begin{document}
\maketitle

\begin{abstract}
We study the excitonic recombination dynamics in an ensemble of (9,4)
semiconducting single-wall carbon nanotubes by high sensitivity time-resolved
photo-luminescence experiments. Measurements from cryogenic to room temperature
allow us to identify two main contributions to the recombination dynamics. The
initial fast decay is temperature independent and is attributed to the presence
of small residual bundles that create external non-radiative relaxation
channels. The slow component shows a strong temperature dependence and is
dominated by non-radiative processes down to 40~K. We propose a quantitative
phenomenological modeling of the variations of the integrated
photoluminescence intensity over the whole temperature range. We show that the
luminescence properties of carbon nanotubes at room temperature are not affected
by the dark/bright excitonic state coupling.

\end{abstract}
\vspace{1cm}

Single-wall carbon nanotubes (SWCNT) are very promising nanoscale materials but,
as expected for objects consisting only of surface atoms, they are highly
sensitive to the coupling with their environment which may dramatically alter
their electronic and optical properties. In fact, the luminescence of
semiconducting SWCNT is one of the most sensitive probes of such
environment-induced effects. Most of the samples do not show any luminescence in
bulk proportions and one has to carefully isolate the nanotube from their
neighbors (and prevent the formation of bundles of nanotubes) in order to
observe radiative recombination across the bandgap \cite{oconnel,
lauretphysicaE}.
However, this effect is only one signature of a more
general change in the electronic properties of the nano-object coupled to its
environment. A better understanding of the nature of the recombination channels in SCWNT is required for any application in photonics or optoelectronics, especially for such a one-dimensional nano-structure where Coulomb interactions are very strong. 
Indeed, it is known from both experimental and theoretical works \cite{heinzscience, maultzsch} that the photoexcited electron-hole pairs form excitons with high binding energy. Due to the symmetry of the nanotubes the lowest energy lying state is expected to be dipole forbidden (dark state) \cite{zhao} which could lead to an intrinsically low quantum yield in agreement with estimates reported in the literature ($Q<10^{-3}$ ) \cite{oconnel, lebedkin, wangPRL}.

Recently, progress in the sample preparation and use of powerful optical techniques gave new insights into the excitonic recombination processes. Time resolved pump-probe measurements have shown that in light-emitting samples where SWCNT are isolated in micelles, the recombination dynamics is at least one order of magnitude slower than in non-emitting samples consisting in SWCNT bundles  \cite{lauretPRB, ostojic}.
Time-resolved photoluminescence (TR-PL) measurements on ensembles of isolated SCWNT \cite{wangPRL, reich} revealed a non-exponential PL decay with a fast component within the first few picoseconds and a long lasting tail of tens of picoseconds. This latter observation led to the conclusion that the intrinsic radiative lifetime may be long in SWCNT.  Recent time-resolved photoluminescence (TR-PL) measurements of individual SWCNT have revealed a monoexponential decay within the experimental sensitivity with a wide dispersion of lifetimes from one tube to another : statistics on (6,4) tubes at 87~K show values spreading from 10 to 180~ps with however a small number of events above 60~ps \cite{hagen}. In that context, ensemble measurements may provide powerful statistical information to identify the recombination mechanisms and their temperature variations.

In this letter, we present high-sensitivity measurements of TR-PL of nanotubes embedded in a gelatin matrix
as a function of the temperature. We propose a simple description of the heterogeneity of the sample which allows us to reproduce the nonexponential temporal response of the sample over three decades at any temperature between 10 and 300~K. The initial fast decay is temperature independent and is attributed to the presence of small residual bundles that create external non-radiative relaxation channels. The slow component shows a previously unresolved linear temperature dependence and is dominated by non-radiative processes down to 40~K. At low temperature, both the integrated PL intensity and lifetime measurements indicate the existence of a regime where carriers are trapped to shallow non-emitting states. From a quantitative phenomenological modeling we deduce an estimate of the dark/bright states splitting in SWCNTs. We emphasize the strikingly weak variations of the PL intensity as a function of temperature and show that the dark/bright excitonic states coupling does not play a key role in the luminescence properties of carbon nanotubes at room temperature.

The sample consists in purified SWCNTs obtained by the HiPCO method and embedded in a gelatin matrix. Following the process described in \cite{oconnel}, we first prepare a suspension of isolated SWCNT by strong sonication in an aqueous solution of SDS (1\% wt). After centrifugation at 200 000 g for 4h the supernatant is collected. In order to obtain a solid sample, we then heat the supernatant at 70~$^o$C and add commercial dehydrated gelatin of low gel point (40~$^o$C). After mixing, a small amount of the solution deposited on a substrate forms a homogeneous gel as it cools down to room temperature. 
The photoluminescence intensity of SWCNT embedded in such a gel is comparable to
the one of the initial suspension (without gelatin), whereas a deposit of
the initial suspension shows a drop of the PL signal of at least one order of
magnitude when the solvant evaporates. We believe that the high hydratation level of the gel preserves the micelle structure based on hydrophilic/hydrophobic competition, in
contrast to the case of an evaporation of the suspension on a substrate where
the reaggregation of nanotubes as bundles is very likely to occur. Moreover, the
SWCNT doped gel is an easy handle solid state composite material that can be cooled down to 4 K and heated back to room temperature  without any
apparent damage, even for tens of cycles.

\begin{figure}
\includegraphics[scale=0.25]{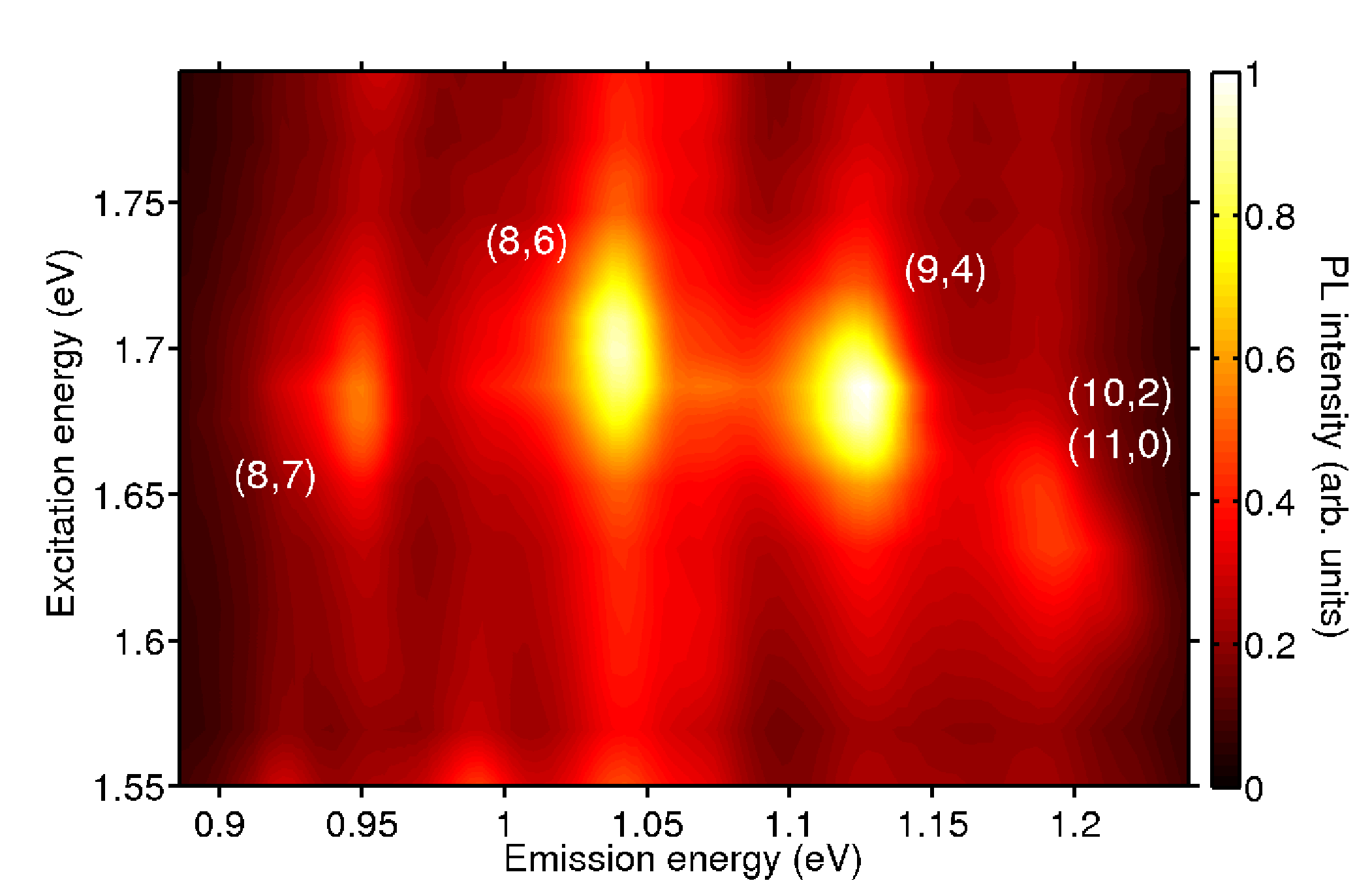}
\caption{Photoluminescence intensity at 10~K of isolated HiPCO SWCNT embedded in
a gelatin matrix plotted as a function of emission and excitation energy. By
comparison to the semi-empirical Kataura plot, PL spots are assigned to a given
SWCNT chirality.} 
\label{fig:PLEmap} 
\end{figure}

The excitation map of the luminescence of the sample at 10~K is displayed in
Fig.~\ref{fig:PLEmap}. The excitation is provided by a cw Ti:sapphire laser and
the detector consists in a InGaAs photodiode. Following the procedure introduced
by Bachilo et al., we assign each peak to a given pair of chiral indices
\cite{bachilo}. For the TR-PL experiments we will focus on the (9,4) chirality
by tuning the excitation energy at 1.7~eV in resonance with the second
 excitonic transition. This
emission line is centered at 1.13 eV with a width of about 40~meV.

Time resolved photoluminescence measurements were performed using 1.4 ps 
pulses from a mode-locked Ti-sapphire laser ($\lambda = 730$ nm $\leftrightarrow$ 1.7 eV). The excitation fluence was kept constant at about 2~$\mu$J.cm$^{-2}$.
The PL signal was spectrally dispersed in a 0.35~m 
monochromator and detected by a synchro-scan streak camera (S1-photocathode) with an mean overall time resolution of 25~ps. TR-PL temporal sections were obtained by spectrally integrating the full line of the (9,4) SWCNT, \textit{i.e.} between 1.08 and 1.165~eV. In order to achieve the best temperature control, the SWCNT doped gel is directly deposited on the cold finger of the cryostat.

\begin{figure}
\includegraphics[angle=0, scale=0.33]{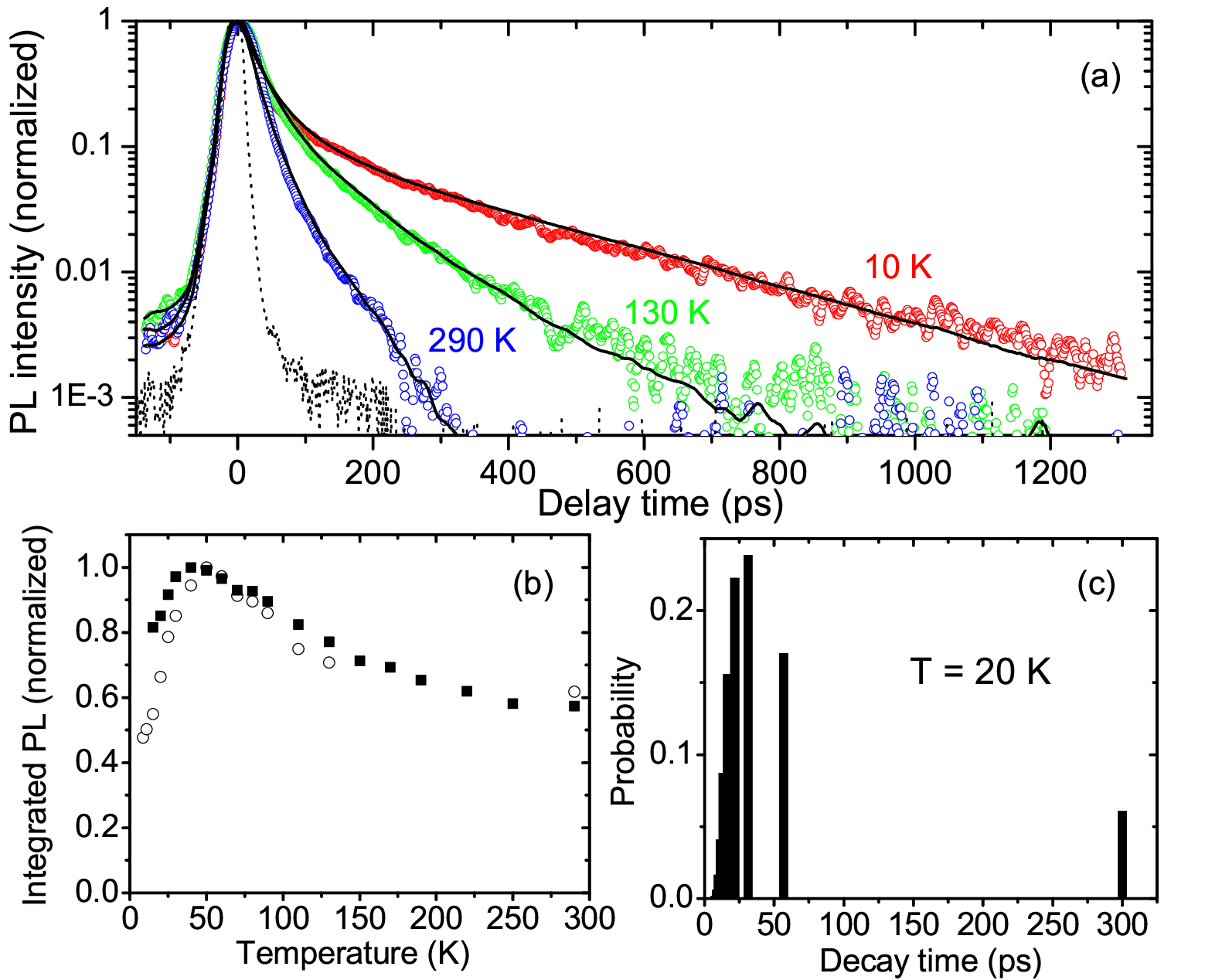}
\caption{(a) Time-resolved photoluminescence of isolated (9,4) SWCNT in a gel on a semi-logarithmic scale for an excitation at 1.7~eV. System response function (dotted line) obtained from the elastic scattering of the laser : the temporal resolution is 26~ps. Normalized photoluminescence transients (open circles) are well reproduced by the convolution (solid lines) of the model $I(t)$ (eq.~\ref{eq2}) and the response function. (b)  Normalized integrated PL signal as a function of the temperature (black dots) ; normalized c.w. PL signal (open circle). (c) Probability distribution of lifetimes at 20~K. } 
\label{fig:PLfit} 
\end{figure}

TR-PL signals are displayed in Fig.~\ref{fig:PLfit}~(a) for different temperatures. The high sensitivity of the setup allows us to detect the signal decay over three decades. We carefully checked that the signal profile remains unchanged when dividing the fluence by up to a factor of 8. Hence we exclude any hidden temperature effect due to laser heating as well as any many body mechanisms such as exciton-exciton annihilation \cite{wang_PRB, ma_PRL, huang_PRL}. Comparison between the temporal integration of these signals and measurements under cw detection (focus mode) shows that we do not miss any hypothetic long living component that could play an important role in the luminescence (Fig.~\ref{fig:PLfit}~(b)). 

At low temperature, the signal shows an initial fast component and after 400~ps a quasi-exponential long living tail. 
This overall non-exponential dynamics is systematically observed for ensemble measurements, whatever the technique \cite{lebedkin, wangPRL, lauretPRB, ostojic}. We believe that it is the signature of the profound inhomogeneity of the sample even within a given chirality. 

In order to go further in the data analysis, we propose a simple model of the inhomogeneity of the sample. Let us first consider the exciton recombination in one single nanotube : in addition to an "internal" decay rate $\gamma_0$ (including both radiative and non radiative contributions), the exciton experiences a small number $j$ of "external" decay channels due to the coupling to the environment, each with a rate $\gamma_{ex}$. Thus the overall rate $\gamma$ varies from one tube to another with the number $j$ of additional channels : $\gamma=\gamma_0 + j \gamma_{ex} $. However the recombination dynamics remains mono-exponential for each tube which is consistent with previous observations of mono-exponential exciton recombination in individual SWCNT \cite{hagen}.
The probability $P_n (j)$ of having $j$ extrinsic channels in a nanotube is assumed to be Poissonian with a parameter $n$ discribing the mean number of external decay channels: $P_n (j)= e^{-n} \frac{n^j}{j!}$.

For a large ensemble of nanotubes,  the TR-PL signal reads :

\begin{eqnarray}
\label{eq1}
I(t) &=& A\sum_{j=0}^{\infty} P_n (j) \exp(-\gamma_0 t - j \gamma_{ex}t)  \\
&=&  A \exp \lbrack-\gamma_0 t - n (1-e^{-\gamma_{ex} t})\rbrack 
\label{eq2}
\end{eqnarray}

where $A$ is the overall amplitude of the signal.

We fit the PL transients with the convolution of $I(t)$ (eq.~(\ref{eq2})) and the system response function. The agreement is excellent for the whole time and temperature ranges as shown in Fig.~\ref{fig:PLfit}~(a). 
We find out that $n$ and $\gamma_{ex}$ are almost independent on the temperature with $n=2.8 \pm0.1$ and $1/\gamma_{ex} = 70 \pm 10$~ps. These values indicate that each tube experiences a small number of relatively efficient additional channels. We attribute this to the presence of remaining small bundles, the external process being a coupling between tubes within the bundle. Indeed, bundles of 2 or 3 nanotubes are hardly distinguishable from single nanotubes in an AFM inspection, especially when surrounded by surfactant molecules. Moreover, such an intertube coupling within a bundle has already been observed by means of time-resolved measurements and has been shown to be temperature independent \cite{lauret}. As a result, this method allows us to extract the response of genuine individual nanotubes (Fig.~\ref{fig:DecayRate}).

From this analysis, we deduce that only 6\% ($P_n(0)$ with $n=2.8$) of the SWCNTs within the sample are effectively individual (and show a decay rate $\gamma_0$). Nevertheless, due to their much lower non radiative decay rate, these tubes provide more than 40\% of the total PL intensity at low temperature and still 15\% at 290~K. The profound inhomogeneity of the SWCNT ensemble is particularly striking when looking at the statistical distribution of exciton lifetimes at a given temperature (Fig.~\ref{fig:PLfit}~(c)). At 20~K, the average decay time ($1/\gamma$) is close to 30~ps whereas the internal decay time ($1/\gamma_0$) is 10 times larger. This histogram of lifetimes is in good agreement with the experimental data of Hagen et al. \cite{hagen} especially concerning the presence of rare events at very large values (10 times the average of the distribution), which is a strong feature of our model.

Following this analysis of the inhomogeneity, we can then estimate the quantum yield $Q_i$ of an individual nanotube ($j=0$ component in eq.(\ref{eq1})) from the average quantum yield $\overline{Q}$ of the sample~: $\frac{\overline{Q}}{Q_i}=\sum_j \frac{\gamma_0  P_n(j)}{\gamma_0+j\gamma_{ex}}$. We find $\frac{Q_i}{\overline{Q}} \simeq 8$ at 40~K. From the values $\overline{Q} \simeq 10^{-3}$ reported in the literature for ensembles \cite{oconnel, lebedkin, wangPRL} (in agreement with our own estimate), we deduce that the quantum yield of an isolated tube can reach one percent. This striking result is in agreement with a recent report of high quantum yield for suspended nanotubes, reaching up to 7\% \cite{lefebvre} and confirms the extreme sensitivity of nanotubes optical properties to their environment.

\begin{figure}
\includegraphics[angle=270, scale=0.31]{Voisin_figure3.eps}
\caption{(a) : Internal decay rate $\gamma_0$ (open circles) of the (9,4) SWCNT as a function of temperature. The solid line is the computed internal decay time from the three level model. (b) : Normalized integrated photoluminescence signal (black squares) from isolated nanotubes (corresponding to the $j=0$ component in eq.~\ref{eq1}) and corresponding fraction in PL measurements under cw excitation (open circles). The solid line is the computed PL intensity from an individual nanotube in the three level model. (Inset) : Schematic of the 3-level model.}.
\label{fig:DecayRate} 
\end{figure}

From the fitting of the temperature dependent data we deduce the evolution of the internal decay rate $\gamma_0$ with temperature (Fig.~\ref{fig:DecayRate}~(a)). At low temperatures (between 10 and 40~K) the internal decay rate remains almost constant when heating up the sample. Then at temperatures higher than 40~K and up to room temperature, the internal decay rate shows a linear increase which has never been reported before. The temperature dependence of the integrated PL intensity (Fig.~\ref{fig:PLfit}~(b)) shows a similar bivalent behavior with a steep increase of the quantum yield when heating up the sample from 10 to 50~K and then a soft decrease between 50~K and room temperature. The peak temperature of 40~K (which corresponds to a typical energy of about 4~meV) thus clearly separates two different regimes in the recombination dynamics.

The temperature evolution of both the integrated PL signal and the recombination time $1/\gamma_0$ in individual nanotubes can be quantitatively reproduced by means of a simple three level model (inset of Fig.~\ref{fig:DecayRate}). The highest level (B) is coupled to the ground state (G) through radiative and non radiative recombination processes with rates $\gamma_R$ and $\gamma_{NR}$ respectively. The radiative decay rate is supposed to be proportional to $T^{-1/2}$ (one dimensional material \cite{andreani}) and much smaller than the non radiative decay rate. The latter is taken proportional to the Bose-Einstein occupation number of a phonon mode $\hbar \omega_p$. The intermediate level (D) is only coupled through non radiative processes to the ground state with the same rate $\gamma_{NR}$ and thus does not contribute to light emission. The two excited states are coupled to each other with rates $\gamma_{\uparrow}$ and $\gamma_{\downarrow}$. 

For each temperature we have performed a numerical computation of the time evolution of the populations and of the integrated PL signal of an isolated nanotube. We find that the coupling rates $\gamma_{\uparrow}$ and $\gamma_{\downarrow}$ between states (B) and (D) have to be much faster than both the radiative and non radiative decays in order to reproduce the experimental data, which means that the populations are almost in thermal equilibrium.
The linear variation of the internal decay rate $\gamma_0$ as a function of the temperature (above 40~K) indicates that the recombination is dominated by phonon assisted processes and is well reproduced with our model by taking a phonon energy of 5~$\pm1$~meV (Fig.~\ref{fig:DecayRate}~(a)). This linear behavior is typical of quasi elastic phonon assisted scattering for which the Bose occupation factor becomes linear for temperatures well above the phonon energy.

We compare the computed PL signal with the PL signal of one isolated nanotube extracted from the ensemble experimental data. This is achieved by selecting the term corresponding to $j=0$ (eq.~\ref{eq1}) in the fit of the experimental TR-PL signal and integrating it over the time. Numerical simulations are in excellent agreement with experiments for an energy splitting between states (B) and (D) of 3.5~$\pm0.5$~meV (Fig.~\ref{fig:DecayRate}~(b)).
We find that the quenching of the PL at low temperature reflects an accumulation
of the carriers in the intermediate state (D). When the sample is heated up, the
occupation probability of the bright state increases while the non radiative
rate remains almost constant, resulting in an increasing PL signal. In the time
domain, the decay rate $\gamma_0$ is almost constant since it is dominated by
the non radiative contribution ($ kT < \hbar  \omega$). Above 40~K the increase
of the non radiative decay rate results in a decrease of the PL signal and an
increase of the overall decay rate.

While our model gives a phenomenological explanation of the data, the nature of
states (B) and (D) is not elucidated. The need of a one-dimensional law
($1/\sqrt{T}$) for the radiative decay rate in order to reproduce the
temperature dependence of the PL intensity suggests that the level (B) is the
delocalized bright excitonic state.
On the other hand, the level (D) can either be the dark state in the one-dimensional exciton picture or a localized shallow defect level. However, the main point is that the splitting between states (B) and (D) is of the order of 3.5~meV (which compares with the lowest theoretical estimates of the one dimensional excitonic splitting \cite{zhao, perebeinos}). This explains why the temperature variations of the PL signal are strikingly weak as compared to the regular behavior in semiconducting nanostructures (for which variations of the PL intensity of several orders of magnitude are commonly observed on the same temperature range \cite{gurioli}). Thus, at room temperature both states are equally populated and the presence of a dark state lying at  lower energy does not play a significant role in the luminescence properties of carbon nanotubes at room temperature.

In summary, we have demonstrated that the inhomogeneity of ensembles of carbon nanotubes and most probably the presence of remaining small bundles is responsible for their non exponential response. We propose a simple modeling to access the internal dynamics of genuine isolated nanotubes. We show that on a large temperature scale above 40~K the variation of the quantum yield is moderate (less than 50\%). This is a consequence of a very small splitting between dark and bright states. This means that the room temperature PL properties (and especially the low average quantum yield) hardly depend on the presence of the dark state. On the other hand we have shown that the inhomogeneity of the sample may hide much larger quantum yields for genuine individual nanotubes. The direct measurement of the quantum yield of one individual nanotube, although challenging, would be of highest interest for future investigation of SWCNT as light emitters. 

The authors are grateful to the whole team of LNMO for technical support and to A. Filoramo and L. Capes for helping in sample preparation. LPA de l'ENS is "Unit\'e Mixte de Recherche associ\'ee
au CNRS (UMR 8551) et aux universit\'es Paris 6 et 7." This work has been done
in the framework of the GDRE n$^{o}$ 2756 'Science and applications of the
nanotubes - NANO-E'. S.B. is funded by a DGA grant.


\end{document}